\documentclass[useAMS,usenatbib]{mn2e}
\usepackage{graphicx}

\newcommand{\aaps}{A\&AS}
\newcommand{\mnras}{MNRAS}
\newcommand{\apj}{ApJ}
\newcommand{\aj}{AJ}
\newcommand{\apjl}{ApJL}
\newcommand{\apjs}{ApJS}

\newcommand{\aap}{A\&A}

\catcode`\@=11
\def\gsim{\ifmmode{\mathrel{\mathpalette\@versim>}}
    \else{$\mathrel{\mathpalette\@versim>}$}\fi}
\def\lsim{\ifmmode{\mathrel{\mathpalette\@versim<}}
    \else{$\mathrel{\mathpalette\@versim<}$}\fi}
\def\@versim#1#2{\lower 2.9truept \vbox{\baselineskip 0pt \lineskip 
    0.5truept \ialign{$\m@th#1\hfil##\hfil$\crcr#2\crcr\sim\crcr}}}
\catcode`\@=12

\def\y1{\hbox{${\rm yr}^{-1}$}}

\title[The hunt for red AGN: a new IR diagnostic]{The hunt for red AGN: a new infrared diagnostic}
\author[D. Fadda and  G. Rodighiero]
{\parbox{\textwidth}{\raggedright Dario Fadda$^{(1)}$\thanks{E-mail: \texttt{fadda@ipac.caltech.edu}},
Giulia Rodighiero$^{(2)}$}\vspace{0.4cm}\\
\parbox{\textwidth}{\raggedright $^{(1)}$IPAC, Caltech 100-22, CA 91125 Pasadena\\
$^{(2)}$Dipartimento di Fisica e Astronomia, Universit\`a di Padova, vicolo dell'Osservatorio 3, I--35122 Padova, Italy.}}

\begin{document}

\date{}

\pagerange{\pageref{firstpage}--\pageref{lastpage}} \pubyear{2014}

\maketitle

\label{firstpage}
\begin{abstract}

We introduce a new infrared diagnostic to separate galaxies on the
basis of their dominant infrared emission: stellar or nuclear.  The
main novelty with respect to existing diagnostics, is the usage of a
broad band encompassing at the same time the 9.7$\mu$m Silicate absorption
feature and one of the adjacent broad PAH (polycyclic aromatic
hydrocarbon) features. 
This provides a robust estimate of the near- to mid-infrared continuum slope and enables a clear distinction among
different classes of galaxies up to a redshift $z\sim 2.5$.
The diagnostic can be applied to a wealth of archival data from the ISO,
Spitzer, and Akari surveys as well as future JWST surveys. Based on
data in the GOODS, Lockman Hole, and North Ecliptic Pole (NEP) fields,
we find out that approximately 70\% active galactic nuclei detected
with X-ray and optical spectroscopy dominate the total mid-infrared
emission.  Finally, we estimate that AGN contribute less than 30\% of
the mid-infrared extragalactic integrated emission.
\end{abstract}


\section{Introduction}
\label{intro}
Star forming regions and active galactic nuclei (AGN) are often
embedded in dust which reprocesses high-energy radiation to
infrared emission.  Infrared surveys are therefore the ideal way to
detect this kind of systems.  The most active star-forming regions are
embedded in dust and contribute most of the extra-galactic integrated
infrared emission (see, e.g. Fadda et al. 2002, Pozzi et al. 2012). On the other
hand, obscured AGN can be detected in the infrared but it is sometimes
difficult to separate them from star-forming dominated systems 
(Feltre et al. 2012,  
Delvecchio et al. 2014). The
same obscured AGN are responsible for the unresolved part of the hard
X-ray cosmic background (e.g. Gilli et al. 2007). 
Even deep X-ray surveys are unable to resolve a substantial part of the
background in the hard X-ray regime. When more than 90\% of the soft
X-ray background has been resolved with recent Chandra and XMM surveys
(Tozzi et al. 2006, Hasinger et al. 2001), only 50\% of the sources in the hard
X-ray band 5-10~keV are currently resolved (Worsley et al. 2005,
Shi et al. 2013).

Waiting for new X-ray detectors in the ultra-hard energy regime
($\sim$30~keV) which will be able to directly see the unobscured AGN,
we can exploit existing icnfrared surveys to uncover this population of
highly extinguished AGN. With the advent of {\sl Spitzer}, several new
techniques have been suggested to select AGN in infrared surveys.
Some of them make use of near- and mid-infrared colours (Lacy et
al. 2004, Stern et al. 2005, Daddi et al. 2007, Hanami et al. 2012,
Donley et al. 2012, Messias et al. 2012, Stern et al. 2012, Juneau et al. 2013, see
Caputi et al. 2014 for a review).
More recently,  {\sl Herschel} far-infrared fluxes have been combined with other wavelengths
to constrain the obscured AGN content in dusty galaxies, through colours and SED fitting
decomposition (e.g. Berta et al. 2013, Kirkpatrick et al. 2013).
However, {\sl Herschel} surveys are not as deep as mid-infrared {\sl Spitzer} surveys.
Despite their success, these techniques suffer of a high degree of contamination
especially at high redshifts and fail to detect Seyfert
type-2 galaxies.  We propose here a new infrared
diagnostic which makes use of a broad mid-infrared band obtained by combining two
filters from the {\sl Spitzer}, {\sl ISO} and {\sl Akari} surveys.
The computed colours  are not affected by the highly variable spectral
features in the mid-infrared wavelength range, allowing a clear distinction among different
classes of galaxies in a large redshift range ($0<z<2.5$).

We apply the diagnostic to {\sl Spitzer} observations in the GOODS
field, {\sl ISO} and {\sl Spitzer} observations in the Lockman Hole,
and {\sl Akari} in the NEP field. Results are compared with X-ray
detections and optical spectroscopy of AGN in the fields. An
estimate of the AGN contribution to the mid-infrared cosmic background
is presented.  Throughout this paper we use $H_0=70$
km~s$^{-1}$~Mpc$^{-1}$, $\Omega_m=0.3$, $\Omega_{\Lambda}=0.7$.

\section{A new Infrared Diagnostic}
\label{diagnostic}
\begin{figure}
\includegraphics[width=8.3cm]{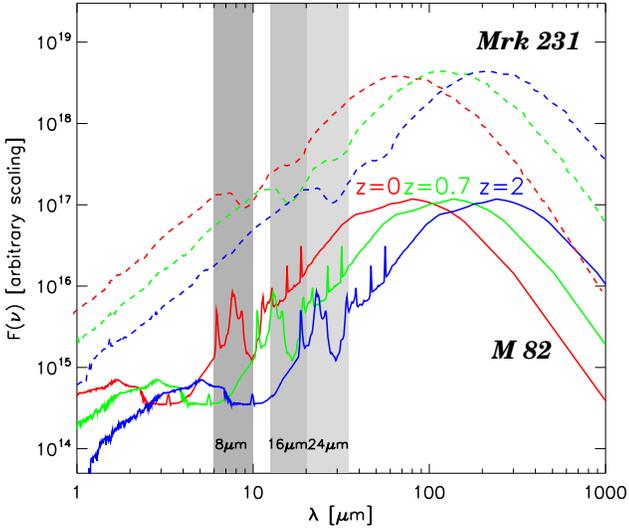}
\label{filters}
\caption{The 8, 16, and 24 $\mu$m  filter coverage of SEDs of a star forming
galaxy (M82, solid lines) and an AGN dominated one (Mrk231, dashed lines) at 3 different redshifts.
}
\end{figure}

Several infrared diagnostics have been proposed to distinguish AGN
from star forming galaxies.  These diagnostics are able to separate
statistically most of the AGN at low redshifts ($z < 1$) while they
become severely contaminated at higher redshifts. Moreover, 
tracks of type~2 Seyfert degenerate on the plots with starburst
galaxies (Rodighiero et al. 2007).
Some otherstudies (see, e.g., Mart{\'{\i}}nez-Sansigre et al. 2005, Daddi et al. 2007,
Fiore et al. 2008, Fadda et al. 2010, Kirkpatrick et al. 2012) use also
24$\mu$m fluxes to select AGN candidates and confirm them through
spectroscopy and/or comparison with X-ray images.

Using the mid-infrared 24$\mu$m flux is, in principle, a very powerful
method to find obscured AGN. The main problem comes from the fact that
the mid-infrared region is full of broad spectral features:
two bands of PAH (polycyclic aromatic hydrocarbon) and the 9.7$\mu$m
Silicate absorption.  The extreme variability of these features among
star forming galaxies and AGN (see Spoon et al. 2007) makes extremely
difficult to use the 24$\mu$m in combination with fluxes at other
wavelengths (e.g. the 8$\mu$m IRAC) as a diagnostic. As shown by Brand
et al. (2006), the 24$\mu$m/8$\mu$m colour cannot be used in a large
range of redshifts to distinguish AGN and star-forming objects.  To
overcome these problems, we propose here to use the
(S$_{24}$+S$_{16}$)/S$_{8.0}$ colour.  As shown in Figure~\ref{filters}, this flux ratio is much less
disturbed by the broad mid-infrared spectral features in a large redshift range ($0<z<2.5$). 
In fact, when one of the filters falls into the
Silicate absorption the other one is on the top of a PAH feature,
compensating for the missing flux. The combination of the two filters
gives therefore a more stable estimate of the continuum and the
proposed flux ratio is, in the redshift interval between 0 and 2.5,
a good estimate of the near- to mid-infrared spectral slope.  Combining this flux ratio with
the IRAC S$_{5.8}$/S$_{3.6}$ colour,  we obtain a very clean diagnostic.
\begin{figure*}
\includegraphics[width=8.7cm]{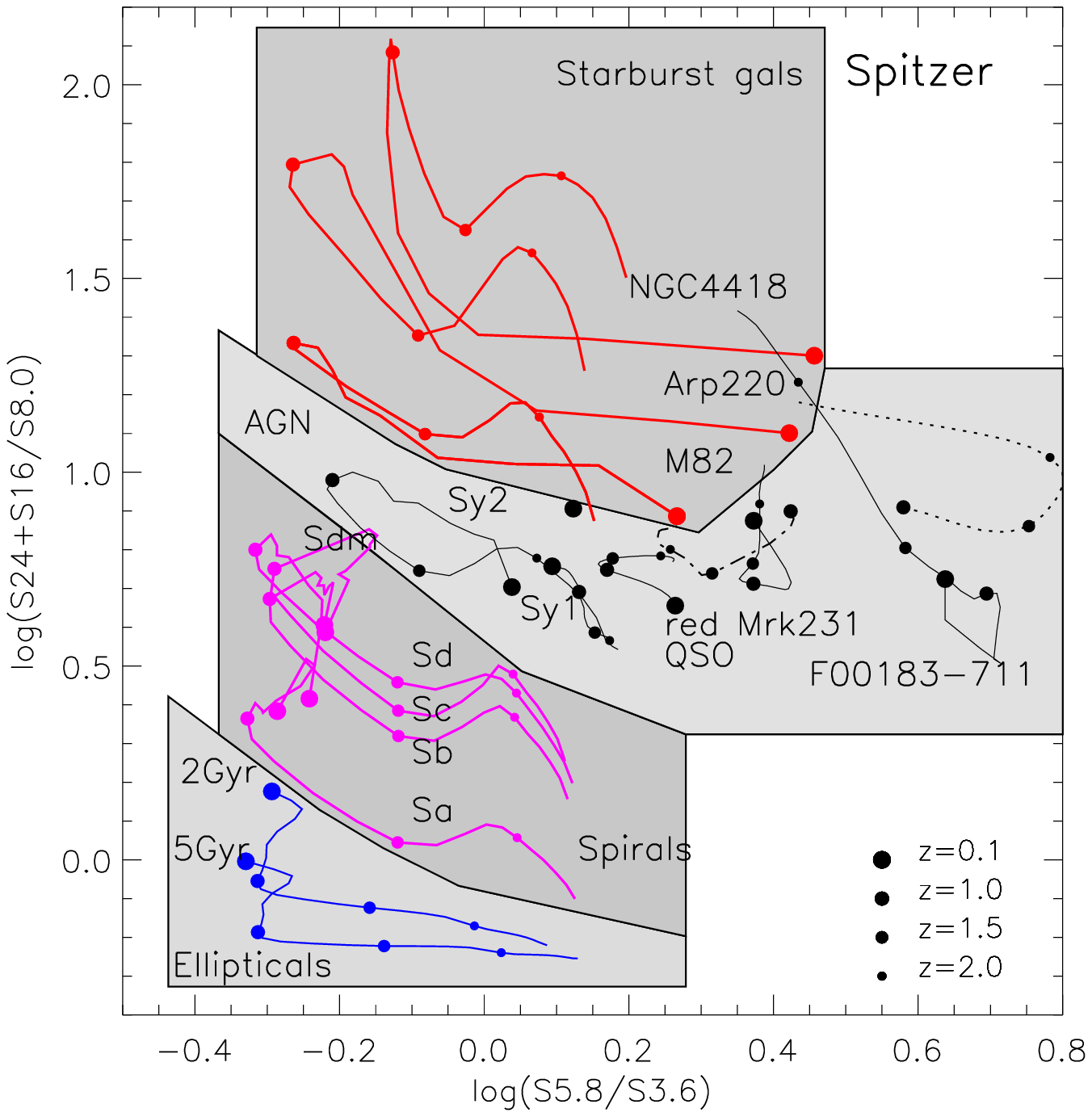}
\includegraphics[width=8.7cm]{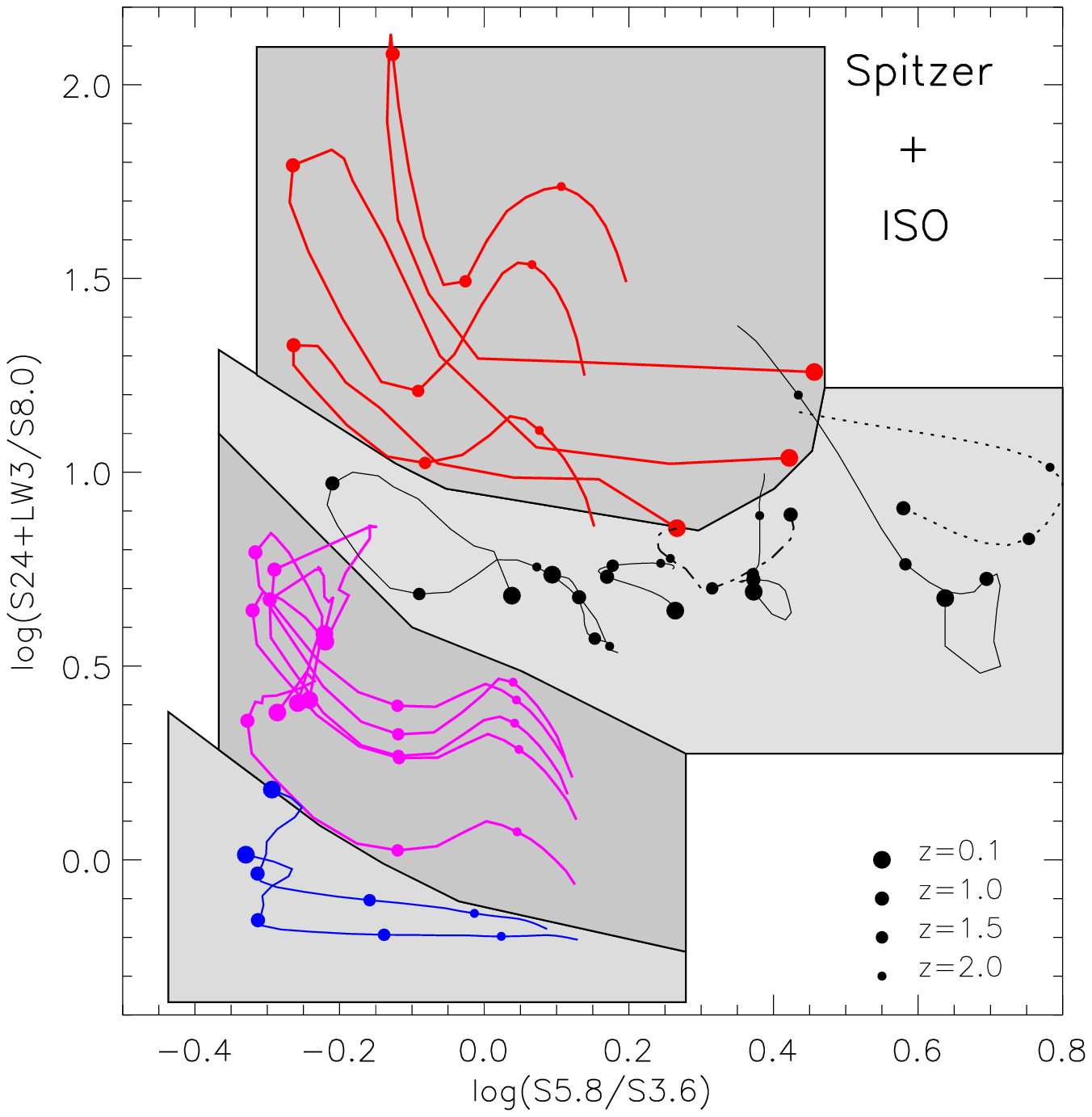}
\includegraphics[width=8.7cm]{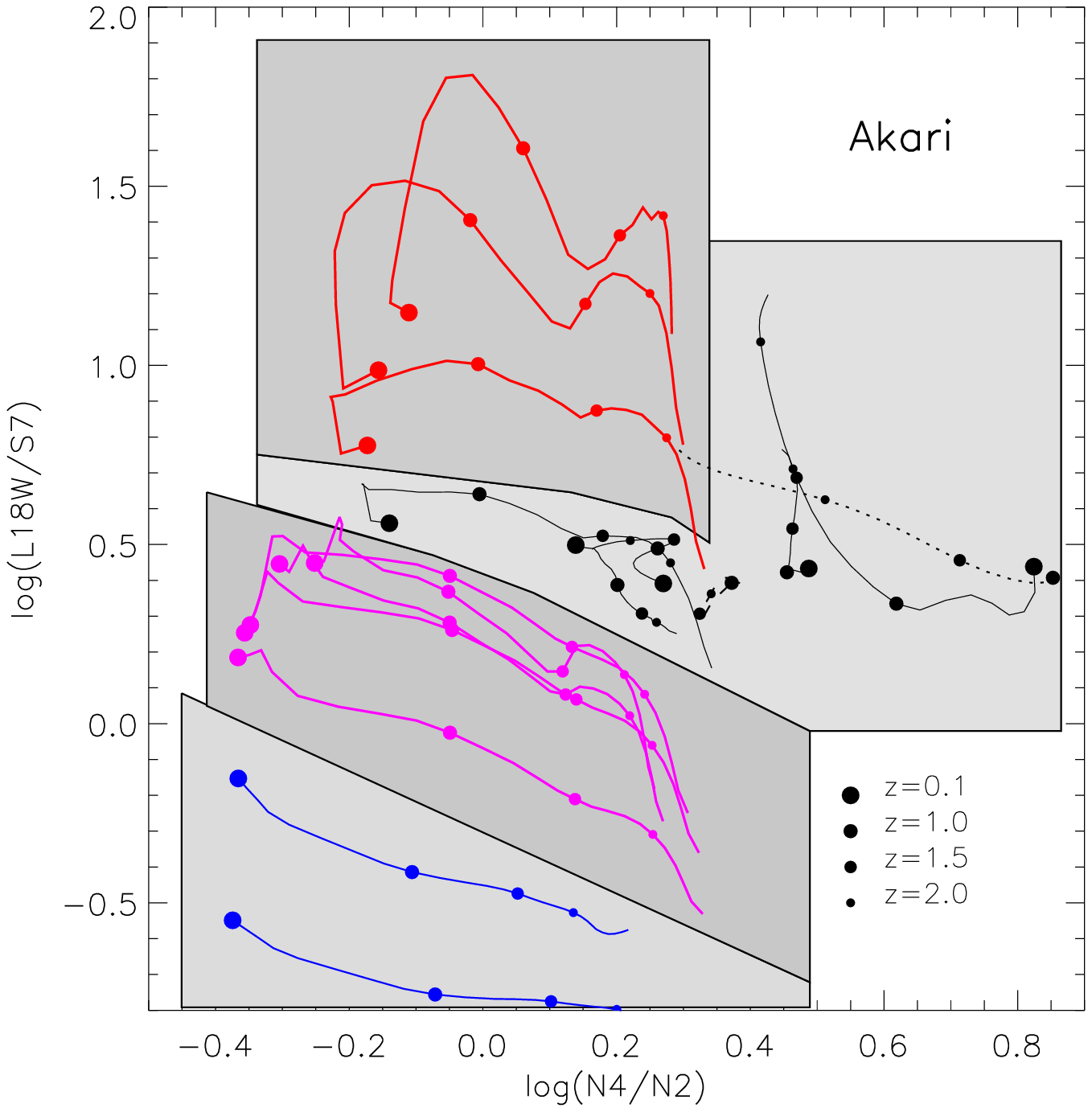}
\includegraphics[width=8.7cm]{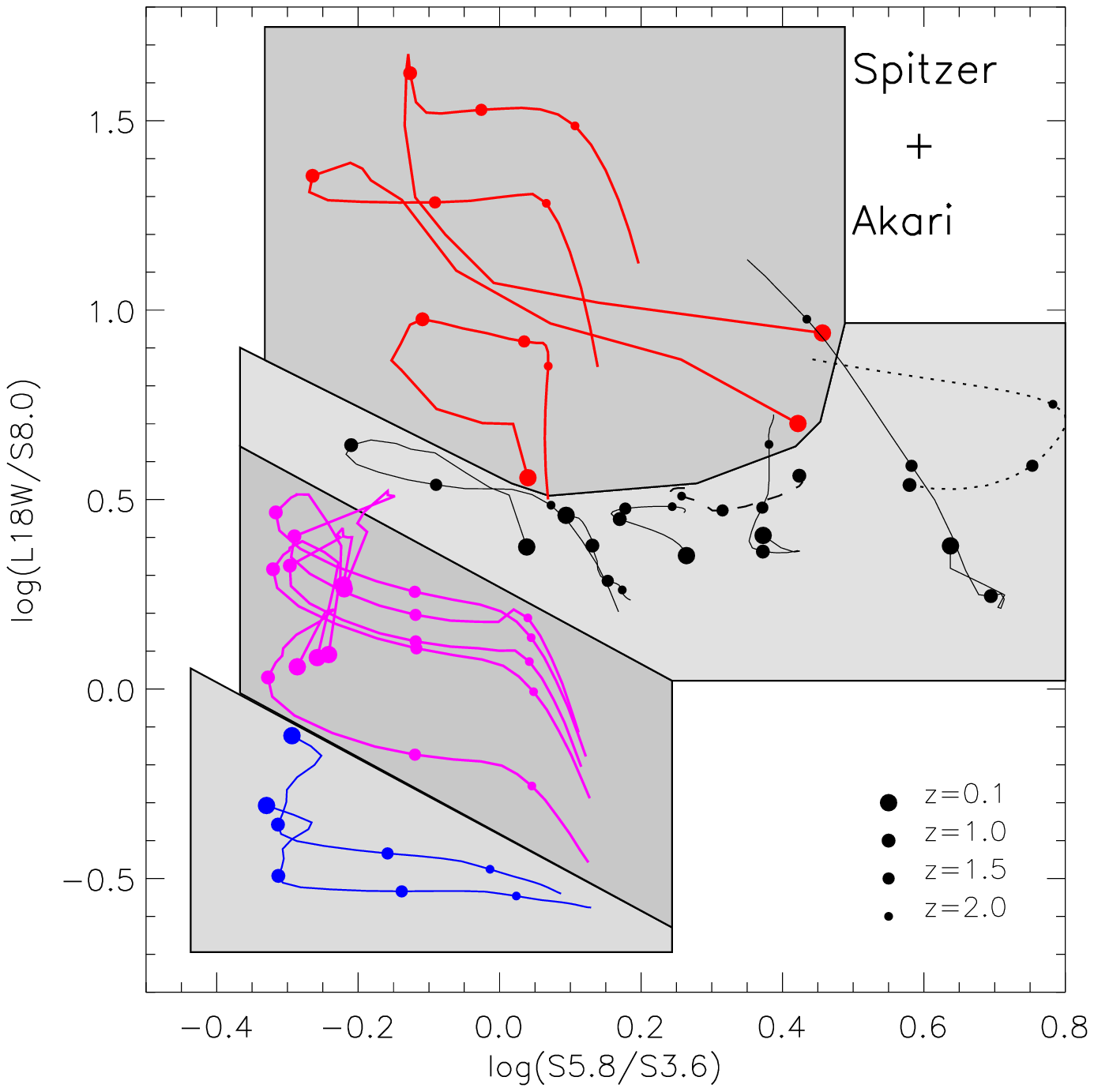}
\caption{The new infrared diagnostic proposed for four different
  combination of filters of {\em ISO}, {\em Spitzer} and {\em Akari}.
  Tracks in the redshift range $0.1<z<2.5$ are of templates from
  Polletta et al. (2007, M82, Sa, Sb, Sc, Sd, Arp220, Sy1, Red QSO,
  Ell 2Gyr, Ell 5Gyr), Spoon et al. (2007, NGC4448 and F00183-7111)
  and Gruppioni et al. (2010, Sy2, Mrk231, and Sdm). 
  Templates of $z>1$ high-$\tau$ AGN (Sajina et al. 2012) and Silicate AGN (Kirkpatrick et al. 2012) are marked with black dotted and dot-dashed lines, respectively.
  We define four regions occupied by different classes of infrared sources.
}  
\label{tracks}
\end{figure*}

\begin{figure*}
\includegraphics[width=17.5cm]{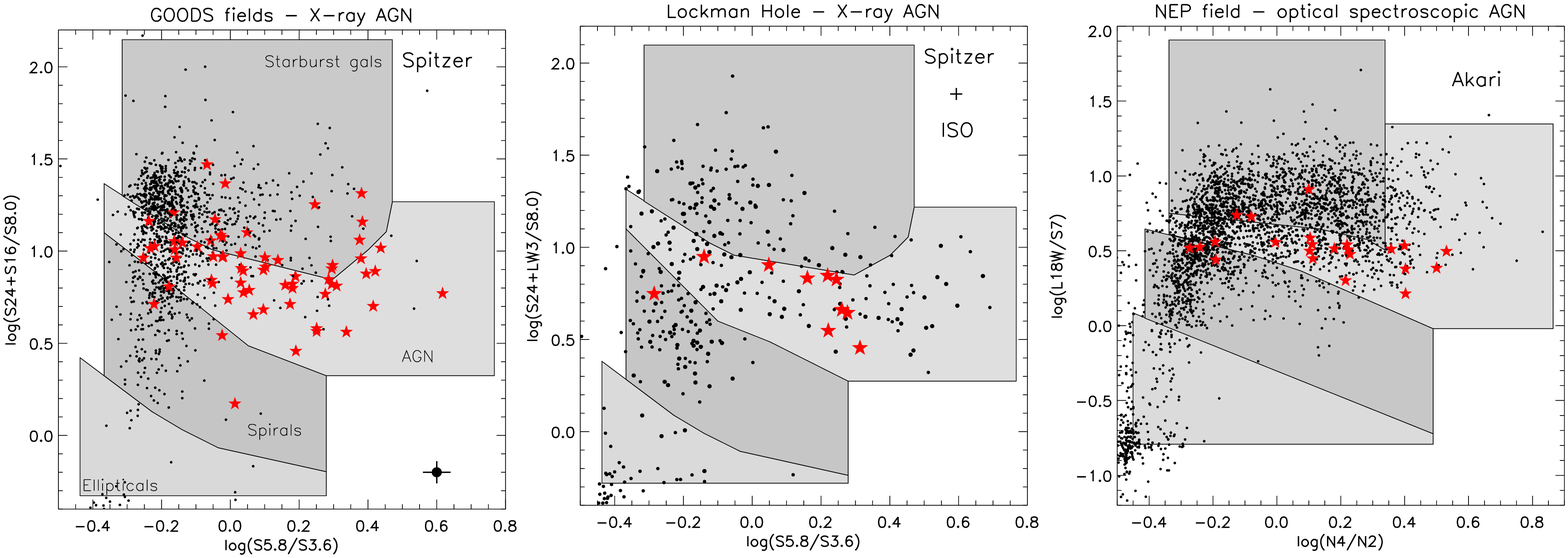}
  \caption{Diagnostic applied to 3 deep fields with AGN shown as red stars:
  {\sl Left:} GOODS and X-ray sources with 
  $L_{2-8 keV} >10^{42.5}L_{\odot}$ (Alexander et al., 2003);
  {\sl Middle:} Lockman Hole and X-ray sources with $L_{2-10 keV} > 10^{42.5}L_{\odot}$;  {\sl Right:}
  NEP field (Murata et al., 2013) and
  sources spectroscopically classified as AGN by Shim et
  al. (2013).  The typical photometric error, assuming 10\% and 5\%
  errors in mid-IR and near-IR fluxes, respectively, is reported in
  the right-bottom corner.
\label{colcol}}
\end{figure*}
Figure~\ref{tracks} shows the tracks of a set of templates 
(see caption for references)
in the range $0.1<z<2.5$ spanning four main classes: ellipticals, spirals, AGN and
starburst galaxies. The four classes are well separated in this
plot, thus proving the strength of this diagnostic to classify
infrared selected galaxies.  Note, in particular, that the Seyfert 2
template (Sy2, in the AGN region) is not degenerate with other
star-forming galaxies (like late spirals), as instead usually happens
for other infrared diagnostics (e.g. Lacy et al. 2005). Rare objects
like the extremely absorbed F00183-711 contaminates the starburst
region only at high redshifts ($z > 2$).
The diagnostic becomes less efficient at redshifts larger than 2.5
when the 1.6$\mu$m stellar bump enters the 8$\mu$m filter affecting the
continuum slope estimate.
By using the template library of Dale et al. (2014), we
verified that objects with emission dominated by AGN (AGN emission greater than 50\% of the total emission) are
selected by our diagnostic.

The same idea can be applied with different sets of filters from other
infrared telescopes, as {\sl ISO} and {\sl Akari}. In the case of {\sl ISO}, the LW3 
band (centered around 15$\mu$m) can be used instead of
the IRS16. For {\sl Akari} (see Figure~\ref{tracks}), one can use the
broad L18W band or the sum of the L24+L15 bands.  Also a combination
of {\sl Akari} and {\sl Spitzer} bands can be used as displayed in
Figure~\ref{tracks}. This is especially useful, since the three
telescopes have publicly accessible archives.

Furthermore, future surveys with the MIRI instrument on board the
James Webb Space Telescope will also provide a set of filters{\footnote{
In particular, to achieve a broad-band as large the 16$\mu$m+24$\mu$m {\sl Spitzer}  filter, one could combine the three MIRI filters F1500W, F1800W and F2550W.}}
suited to apply the proposed diagnostic. 

\section{Data}
\label{data}

In order to apply the proposed diagnostic, we used a set of publicly
available {\em Spitzer},  {\em ISO}, and {\em Akari} data in four different deep
fields: GOODS North, GOODS South, Lockman Hole, and North Ecliptic Pole (NEP).  The first three
fields have been observed with Legacy (SWIRE, Londsale et al. 2003) and
GTO programs with {\em Spitzer} in all the infrared bands of IRAC and
MIPS.  IRAC and MIPS highly processed images are available from the
{\sl Spitzer} archive.  {\em ISO} observations of the Lockman Hole
with ISOCAM in the LW3 band have been published by the authors (Fadda
et al. 2004, Rodighiero et al. 2004).  Moreover, the two GOODS fields
have been observed with IRS onboard {\em Spitzer} at 16$\mu$m. These
data are now publicly available from the {\em Spitzer} archive and we
have retrieved and processed them.

To reduce the 16$\mu$m data we started from the basic calibrated data (BCD)
available from the {\em Spitzer} archive. Stacking all the BCDs, we obtained
a superflat which has been applied to the single BCDs. Since we are only
interested in the point sources in the field, we subtracted also the 
median background from each BCD before mosaicking them in a single image.
We used the {\em Spitzer} software MOPEX to obtain a mosaic with a pixel 
half the size of the original ones using the redundancy of the observations
to reject outliers.

For the source extraction we used a PSF fitting method (Starfinder,
Diolaiti et al. 2000) for the 16$\mu$m and MIPS 24$\mu$m data and
aperture photometry with SExtractor (Bertin \& Arnout, 1996) for the
IRAC images. In the case of IRAC images, in fact, PSF fitting
techniques are hampered by the undersampling of the PSF.  A critical
step to apply the colour diagnostic proposed in the paper is to
accurately compute aperture corrections for the different filters.  As
shown by Fadda et al. (2006), the 24$\mu$m theoretical PSF obtained
using
STinyTim\footnote{irsa.ipac.caltech.edu/data/SPITZER/docs/dataanalysistools/
tools/contributed/general/stinytim}
accurately reproduces the real 24$\mu$m PSF. The same is true for the
16$\mu$m array.  So, in these cases, we fitted a PSF inside the
aperture of 9.45 arcsec and 14.0 arcsec radii for the 16$\mu$m and
24$\mu$m bands, respectively.  The aperture correction computed was of
1.1626 and 1.158 for the 16$\mu$m and 24$\mu$m bands, respectively.

In the case of IRAC, the real PSF is significantly different from that
produced with STinyTim and the model cannot be used to compute
aperture corrections. So, we used the values computed with a set of
stars in the SWIRE fields (see SWIRE data release
report\footnote{swire.ipac.caltech.edu/swire/astronomers/publications/
SWIRE2\_doc\_083105.pdf},
page 31).  Using 3.8 arcsec diameter apertures, we applied corrections
of 1.35, 1.39, 1.65, and 1.84 for the 3.6, 4.5, 5.8, and 8.0 $\mu$m
IRAC channels, respectively.

The final catalogs have been matched starting from the 16$\mu$m (or
15$\mu$m) catalog using a matching radius of 2 arcsec. The 16$\mu$m and
24$\mu$m filters are expected to be sensitive to different redshift
regions because the 9.7$\mu$m Si absorption affects them at different
redshifts. In our cases, because of the depth of the IRAC and MIPS
images, only a few 16$\mu$m sources do not have MIPS counterparts.
Above 0.1 mJy, the percentage of 16$\mu$m sources without MIPS counterparts
is negligible (less than 10\%) for all the surveys here considered. 
For fluxes fainter than 0.1 mJy, the percentage is slightly higher: 
in particular, for GOODS-North the percentage is still around 10\%
down to 0.06mJy, while for the UDF the percentage gets worse in the
fainter flux bin considered in the following discussion (30-40\% in
the range 0.04-0.06 mJy).

For what concerns the NEP {\em Akari} field, we rely completely  on published photometry
(Murata et al. 2013).

\section{Comparison with X-ray and spectroscopic selections}
\label{application}

X-ray surveys are one of the cleanest methods to select AGN since they
probe directly their high energy emission. However, the situation is
more complex when dusty tori obscure the X-ray emission and
reprocess it into infrared emission. For these reasons, the AGN
detections through X-ray and infrared are somewhat complementary.
However, when using infrared bands to detect AGN, we have to remember that
  dusty tori are not usually the main contributors to the total infrared emission
  except for the very bright end of the infrared population (see, i.e., Brand et al. 2006).

To test how many of the infrared sources classified as AGN based on
their X-ray emission are classified as AGN by the our diagnostic, we
considered three fields which have been observed with {\sl Spitzer}
and {\sl ISO} in the infrared and with {\sl Chandra} (Alexander et
al. 2003) and {\sl XMM} (Hasinger et al. 2001) in the X-ray.  We can
apply the diagnostic using only {\sl Spitzer} data to the two GOODS
fields which have been observed with IRAC and MIPS with the GOODS
legacy survey.  To compute the X-ray luminosities, we used a
compilation of redshifts for the GOODS
fields\footnote{www.eso.org/sci/activities/garching/projects/goods/MasterSpectroscopy}
and redshifts from Lehmann et al. (2001) for the Lockman Hole. As
shown in Figures~\ref{colcol}, most of the sources with high X-ray
fluxes (i.e. $L_{2-10 keV} >10^{42.5}L_{\odot}$), that are probably
AGN, fall in the AGN locus of our empirical diagnostic. Taking into
account photometric errors, (70 $\pm$ 10)\% of the GOODS X-ray sources
fall in the AGN locus. The percentage is approximately 90\% in the
case of the Lockman Hole. As already pointed out, the diagnostic is
not expected to be 100\% accurate because of the mixed source of
infrared emission in galaxies. Especially for distant AGN, the host
galaxy can produce enough stars to mask the AGN behaviour in the low
resolution {\em Spitzer} images. On the contrary, in the case of X-ray
observations, the AGN dominates the total emission even in the case of
low resolution images.

In the NEP field, where a complete {\sl Akari} photometric catalog is
available (Murata et al. 2013), we can test our diagnostic against a
selection of AGN sources classified on the basis of their optical
emission lines (Shim et al. 2013). 70$^{+10}_{-5}$\% of the optically
detected AGN are classified as infrared sources with emission dominated
by the AGN by our diagnostic.

Our findings are compatible with the study of Polletta et al. (2007)
who find that $\sim$~80\% of X-ray selected AGN are identified as AGN
using their infrared SEDs when the redshift of the source is known, 
and with Donley et al. (2012) who were able to recover 75\% of the
  X-ray sources brighter than $L_{2-10 keV}=10^{44} erg/s$ with their  IRAC color selection technique.
On the basis of these comparisons, we argue that the proposed
diagnostic can be safely used to obtain an estimate of the AGN
contribution to the mid-IR background.  A further validation of the
diagnostic with SED fitting decomposition (including a dusty torus
component) is deferred to a forthcoming paper (Baronchelli et al., in
preparation).

\begin{figure}
\includegraphics[width=8.5cm]{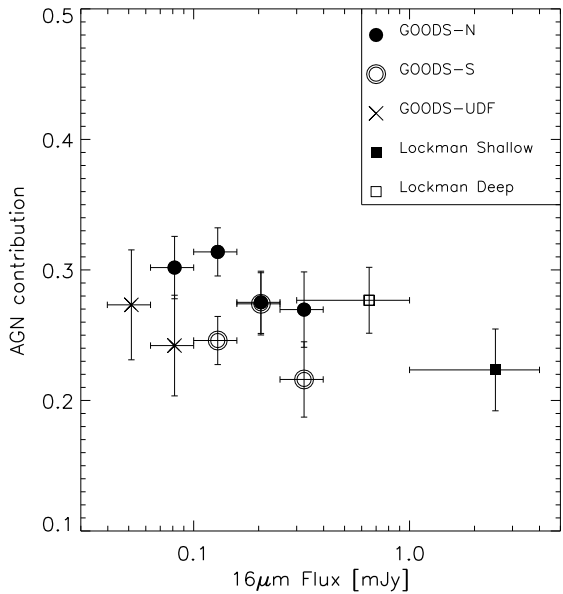}
\caption{AGN contribution to the cosmic infrared background at 16$\mu$m
as a function of the flux using surveys in three different fields.
\label{contrib}}
\end{figure}

\section{AGN contribution to the mid-IR background}
\label{background}

To compute the AGN contribution to the mid-IR background, we considered
the sum of the fluxes of the sources falling in the AGN loci defined in
Figure~\ref{tracks}.  The total extragalactic flux has been computed by
adding the fluxes of all the sources which are not stars. 
In our diagnostic, stars fall at the bottom of the colour diagram and can
be easily rejected. Figure~\ref{contrib} reports the fraction of AGN flux
from the different fields in several flux bins. Poissonian error bars
are reported. Due to the low number of sources, errors are still rather
large but all the measurements are consistent and give an estimate
of the AGN contribution between 20\% and 30\%. The fraction is larger than
the estimate done in the Lockman Hole using only X-ray identified AGN
by Fadda et al. (2002) and is closer to more recent estimates based on
{\em Spitzer} 24$\mu$m data (Choi et al. 2011).
A better coverage of the mid-infrared region with future JWST surveys
will allow the selection of large samples of highly obscured AGN.

\section*{Acknowledgments}
We thank the referee for his/her suggestions that improved the clarity of the paper.
GR acknowledges support from the University of Padova from ASI
(Herschel Science Contract I/005/07/0).  This work is based in part on
observations made with {\em Spitzer}, a space telescope operated by
the Jet Propulsion Laboratory, California Institute of Technology
under a contract with NASA. Support for this work was provided by NASA
through an award issued by JPL/Caltech.
We thank Alberto Franceschini for useful comments.

\end{document}